\documentclass[prl,twocolumn,showpacs,preprintnumbers,amsmath,amssymb]{revtex4}

\usepackage{graphicx}% Include figure files
\usepackage{dcolumn}% Align table columns on decimal point
\usepackage{bm}% bold math

\begin{document}

\newcommand{\spinup}{\protect{$ \left|\uparrow \right\rangle$}}
\newcommand{\spindown}{\protect{$ \left|\downarrow \right\rangle$}}
\newcommand{\upeq}{\protect{\left | \uparrow \right\rangle}}
\newcommand{\downeq}{\protect{\left | \downarrow \right\rangle}}
\newcommand{\upxeq}{\protect{\left | \uparrow_x \right\rangle}}
\newcommand{\downxeq}{\protect{\left | \downarrow_x \right\rangle}}

\newcommand{\uu}{\protect{$ \left|\uparrow\uparrow\right\rangle$}}
\newcommand{\dd}{\protect{$ \left|\downarrow\downarrow \right\rangle$}}
\newcommand{\ud}{\protect{$ \left|\uparrow\downarrow \right\rangle$}}
\newcommand{\du}{\protect{$ \left|\downarrow\uparrow \right\rangle$}}
\newcommand{\uue}{\protect{\left|\uparrow\uparrow\right\rangle}}
\newcommand{\dde}{\protect{\left|\downarrow\downarrow \right\rangle}}
\newcommand{\ude}{\protect{\left|\uparrow\downarrow \right\rangle}}
\newcommand{\due}{\protect{\left|\downarrow\uparrow \right\rangle}}

\title{Entanglement of Trapped-Ion Clock States}

\author{P.~C. Haljan, P.~J. Lee, K.-A. Brickman, M. Acton, L. Deslauriers, C. Monroe}
%\email{haljan@umich.edu}
\affiliation{FOCUS Center and University of Michigan Department of Physics}

\date{13 Aug 2005}% It is always \today, today,

\begin{abstract}
A M{\o}lmer-S{\o}rensen entangling gate is realized for pairs of trapped
$^{111}$Cd$^+$ ions using magnetic-field insensitive ``clock" states and an
implementation offering reduced sensitivity to optical phase drifts. The gate
is used to generate the complete set of four entangled states, which are
reconstructed and evaluated with quantum-state tomography. An average
target-state fidelity of 0.79 is achieved, limited by available laser power
and technical noise. The tomographic reconstruction of entangled states
demonstrates universal quantum control of two ion-qubits, which through
multiplexing can provide a route to scalable architectures for trapped-ion
quantum computing.
\end{abstract}


\maketitle

Entangled states such as the famous EPR-Bohm states \cite{EPR,BohmQM} have
long been of interest in the interpretation of quantum mechanics
\cite{BellSpeakable}; however, their generation has become a rapidly growing
field with the recognition of entanglement as a powerful resource for quantum
information processing\cite{NielsenChuang}. Laser-addressed trapped ions
with qubits embedded in long-lived internal hyperfine levels hold
significant advantages for quantum information applications
\cite{wineland98b,blinov2004a}. A critical issue is the robust generation of
scalable entanglement. In the context of trapped ions this is reduced to the
problem of two-qubit entanglement, as plausible multiplexing schemes have
been proposed to create a scalable architecture for a quantum
processor\cite{Kielpinski2002a,Rowe2002a}.

Trapped-ion entangling gates mediated by phonons of the collective ion motion
are susceptible to various forms of noise - qubit and motional decoherence,
impure initial conditions, and technical issues associated with the optical
Raman lasers driving the gate\cite{wineland98b}. Robust schemes for gates
based on spin-dependent forces have been proposed
\cite{molmer99,solano1999a,milburn2000a,Garcia-Ripoll2003a} and
experimentally implemented \cite{sackett00,leibfried2003a} that, for
example, relax the purity requirement on the initial motional state of the
ions. Here we report the realization of one such entangling gate for pairs of
trapped $^{111}$Cd$^+$ ions that uses an advantageous
implementation\cite{haljan2005a,Lee2005a} of the M{\o}lmer-S{\o}rensen (MS)
scheme\cite{molmer99,sackett00}. The implementation reduces sensitivity to
optical phase drifts through an appropriate Raman beam setup and reduces
sensitivity to magnetic field fluctuations through the use of magnetic-field
insensitive clock states\cite{Langer2005a}.

Quantum state tomography
\cite{tomoreview,james2001a,roos2004a,altepeter2004a} is used to
characterize the gate performance for the creation of all four entangled
Bell-like states. Previous applications of quantum state tomography with
ions include the reconstruction of non-classical states of motion
\cite{Meekhof1996a,leibfried96,leibfried2003b} as well as entangled states
of optical ion-qubits composed of electronic levels \cite{roos2004a}. Here
we present the first such implementation for hyperfine qubits, in the process
demonstrating universal quantum control of two clock-state ion-qubits.

%%%%%%%%%%%%%%%%%%%%%%%%%%%%%%%%%%%%%%%%%%%%%%%%%%%%%%%%%%%%%%%%%%%%%%%%%%
%figure MSgate
\begin{figure}[b]
\begin{center}
\includegraphics[width=\linewidth,clip]{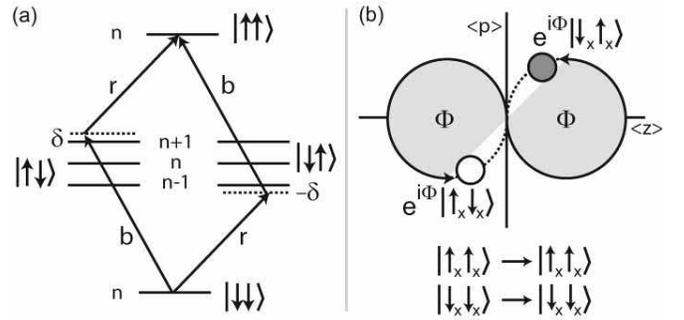}
\end{center}
\caption{Two views of the M{\o}lmer-S{\o}rensen
$\hat{\sigma}_x\otimes\hat{\sigma}_x$ entangling gate for two ions in (a)
energy space \cite{molmer99} and (b) motional phase space
\cite{leibfried2003a} for the gate-diagonal spin basis. The quantum number
$n$ and phase-space co-ordinates describe a given collective motional mode.
Red and blue Raman sideband couplings are labeled by $r$ and $b$ and have
detuning $\delta_b\!=\!\delta\!=\!-\delta_r$. For a closed phase-space
trajectory, the phase $\Phi$ depends only on the area enclosed.}
\label{fig:MSgate}
\end{figure}
%%%%%%%%%%%%%%%%%%%%%%%%%%%%%%%%%%%%%%%%%%%%%%%%%%%%%%%%%%%%%%%%%%%%%%%%%

The MS gate for two trapped ions is based on optical Raman couplings to the
first vibrational sidebands of the ions' collective motion, assumed along
the $z$-axis. The ions are equally illuminated by a bichromatic Raman field
inducing simultaneous red and blue sideband interactions
\cite{leibfried2003b} that couple each ion's spin to the vibrational levels
$\{|n\rangle\}$ of a single collective mode of motion - stretch or center of
mass [Fig.~\ref{fig:MSgate}(a)]. With the Raman fields far detuned from the
sidebands, negligible direct coupling occurs; however, the fields can
combine to provide a resonant two-step coupling between for example \dd~and
\uu, generating the entangled state $1/\!\sqrt{2}(\dde + i\uue)$
[Fig.~\ref{fig:MSgate}(a)]. In general, the bichromatic field provides an
entangling gate based on a nonlinear two-qubit interaction such as
$H\!=\!-\hbar\tilde{\Omega}(\hat{\sigma}_x\otimes\hat{\sigma}_x)/2$, written
in terms of Pauli operators. The coupling strength $\tilde{\Omega}$, which
in the Lamb-Dicke limit is independent of the initial value of
$\left|n\right\rangle$, is given by $(\eta\Omega)^2/\delta$ where $\Omega$
is the carrier Rabi frequency  and $\delta$ the detuning. The Lamb-Dicke
parameter $\eta=k z_o$ for the motional mode of interest is characterized by
Raman wavevector difference $k$ along the $z$-axis of motion and zero-point
wavepacket size $z_o\!=\!\sqrt{\hbar/2 M \omega}$, where $\omega$ and $M$ are
the frequency and total mass of the excitation respectively.

Reducing the detuning $\delta$ accelerates the gate speed at the expense of
populating intermediate motional states \cite{Sorensen2000a}. In this
situation it is more natural to view the bichromatic field as generating a
spin-dependent force constructed from red and blue sideband couplings with
balanced Rabi frequencies and detunings
\cite{bichromatic,haljan2005a,Lee2005a}. The resulting interaction on each
ion is equivalent to $H\!\sim\!\hat{\sigma}_xzF_o\sin\omega_d t$ describing a
$\hat{\sigma}_x$-dependent force near resonance
($\omega_d\!=\!\omega+\delta$) with strength $F_o z_o$=$\eta\Omega$. The
total Hamiltonian is the sum of interactions on each ion. For a force
resonant with the stretch mode and acting in-phase on the two ions, the
time-evolution operator can be expressed in the $\hat{\sigma}_x$-diagonal
gate basis as a spin-dependent displacement as follows:
\begin{equation} \label{Ugate}
\begin{array}{l}
\hat{U}(t)=\left|\uparrow_x\uparrow_x\rangle\langle\uparrow_x\uparrow_x\right|
+
\left|\downarrow_x\downarrow_x\rangle\langle\downarrow_x\downarrow_x\right|
\\
+
e^{-i\Phi}\hat{\mathcal{D}}(\alpha)\left|\uparrow_x\downarrow_x\rangle\langle\uparrow_x\downarrow_x\right|
+
e^{-i\Phi}\hat{\mathcal{D}}(-\alpha)\left|\downarrow_x\uparrow_x\rangle\langle\downarrow_x\uparrow_x\right|
\end{array}
\end{equation}
\noindent where $\hat{\mathcal{D}(\alpha)}$ is the displacement operator in
the phase space of the driven normal mode [Figure 1(b)]. The value of the
displacement is $\alpha(t,\delta)\!=\!\alpha_o(1-e^{-i\delta t})$ and the
corresponding phase accumulated over the trajectory is
$\Phi(t,\delta)\!=\!\alpha_o^2(\delta t-\sin\delta t)$ in terms of the
parameter $\alpha_o$=$\eta\Omega/\delta$. In general, the spin-dependent
displacement entangles the spin and motional degrees of freedom; however,
for a closed trajectory ($\delta t\!=\!2\pi m$, $m$ an integer), the spin and
motion disentangle leaving only a spin-dependent geometric phase
$\Phi_g\!=\!2\pi m (\eta\Omega/\delta)^2$ applied to the gate basis. A
maximally entangling phase gate is constructed from a geometric phase of
$\pi/2$. We achieve this in the fastest time possible with $m\!=\!1$
requiring detuning $\delta\!=\!2\eta\Omega$ and gate time
$\tau_g\!=\!2\pi/\delta$. Expressed in the computational basis, the MS gate
structure, although slightly more complicated, makes the entangling action
manifest:
\begin{equation}\label{eqn:MSgate2}
 \begin{array}{l}
     \left|\uparrow\uparrow\right\rangle
     \rightarrow \Psi_1=\frac{1}{\sqrt{2}}\left(\left|\uparrow\uparrow\right\rangle +  i e^{i\phi_e}\left|\downarrow\downarrow\right\rangle\right) \\
     \left|\downarrow\downarrow\right\rangle \rightarrow \Psi_2=\frac{1}{\sqrt{2}}\left(\left|\downarrow\downarrow\right\rangle +
     i e^{-i\phi_e}\left|\uparrow\uparrow\right\rangle\right) \\
     \left|\uparrow\downarrow\right\rangle \rightarrow \Psi_3=\frac{1}{\sqrt{2}}\left(\left|\uparrow\downarrow\right\rangle +
     i e^{i\phi_o}\left|\downarrow\uparrow\right\rangle\right)  \\
     \left|\downarrow\uparrow\right\rangle \rightarrow \Psi_4=\frac{1}{\sqrt{2}}\left(\left|\downarrow\uparrow\right\rangle +
     i e^{-i\phi_o}\left|\uparrow\downarrow\right\rangle\right)
 \end{array}
\end{equation}
\noindent The phases $\phi_o$ and $\phi_e$ have been included in the even
and odd parity states, $\Psi_{1,2}$ and $\Psi_{3,4}$ respectively, to account
for the effect of both ac Stark shifts and Raman laser coherences, the latter
modifying the spin dependence of the gate \cite{Lee2005a,phasenote}. For
$\phi_e\!=\!\phi_o\!=\!0$, the gate's action reduces to that of a
$\hat{\sigma}_x\otimes\hat{\sigma}_x$ coupling.

Our qubit resides in the hyperfine clock states
$\upeq$=$|F$=$0,m_F$=$0\rangle$ and $\downeq$=$|F$=$1,m_F$=$0\rangle$ of a
$^{111}$Cd$^+$ ion with frequency separation $\omega_{h\!f}/2\pi$=14.53GHz
and second-order Zeeman shift 600Hz$/$G$^2$ near zero magnetic field. A pair
of ions, confined in a three-layer linear Paul trap \cite{Deslauriers2004a}
is aligned along the weak \textit{z}-axis. The ions' secular harmonic motion
in the \textit{z}-direction is characterized by center-of-mass and stretch
normal modes with frequencies $\omega_c$=2.05MHz and
$\omega_s$=$\sqrt{3}\omega_c$. All requirements for arbitrary two-qubit
control are implemented as follows. Both modes of motion are initialized to
near their ground state ($\bar{n}_{c}\!\sim\!0.4,\bar{n}_s\!\sim\!0.2$) with
60 pulses of Raman sideband cooling \cite{Deslauriers2004a}. Due to the
simplicity of the hyperfine structure (nuclear spin $I\!=\!1/2$), the qubits
are directly initialized with optical pumping to \spinup. Following coherent
operations the qubits can be read out with high fidelity using a
photomultiplier tube (PMT) allowing \uu, \dd~and \{\ud, \du\} to be
distinguished by virtue of state-dependent fluorescence \cite{blinov2004a}.
Unambiguous readout of all four qubit states, in particular \ud~and \du, is
achieved with an intensified CCD camera that can independently and
simultaneously image the fluorescence collected from each ion-qubit. A
camera detection fidelity of 97\% is achieved in 15ms limited by readout
electronics.

Coherent single qubit operations are achieved through a combination of
applied microwave fields and ion-selective ac Stark shifts. Resonant
microwaves provide simultaneous Rabi flopping of both qubits with a Rabi
frequency of 56kHz. Arbitrary \textit{independent} qubit rotations are
achieved by combining microwave operations with pulses of an off-resonant
laser beam 200GHz detuned from resonance. The laser beam with moderate waist
($\lesssim$10$\mu$m compared with the 2.5$\mu$m ion spacing) is aligned to
be off-center with respect to the two ions, giving rise to an intensity
gradient and differential ac Stark shift between the two ions. A 10$\mu$s
exposure results in a phase shift difference between the two qubits of
approximately $\pi$/2.

The MS entangling gate is achieved using a pair of Raman laser beams 200GHz
detuned from optical resonance. An electrooptic modulator(EOM) at microwave
frequency together with an acousto-optic modulator in each Raman beam provide
the required bichromatic Raman beatnote \cite{lee2003a}. The collective
stretch mode is chosen for gate implementation due to the significantly
suppressed heating rate \cite{King1998a}. The stretch sideband Rabi
frequency is typically 6kHz. The wavevectors of the red and blue Raman
fields are arranged in a counter-propagating geometry so that the spin
coherence of the MS gate, included in $\phi_e$, is insensitive to optical
phase drifts between the Raman beams \cite{haljan2005a,Lee2005a}. Although
this setup requires an accurate ion spacing to maximize gate speed, the
stability of the spin coherence is crucial for keeping the MS gate
synchronized with the microwave fields during tomography. A noise eater,
stabilizing the Raman beam power, is used to suppress the effect of
fluctuating ac Stark shifts (where the average qubit shift is 75kHz compared
with 13kHz gate speed as discussed below).

%%%%%%%%%%%%%%%%%%%%%%%%%%%%%%%%%%%%%%%%%%%%%%%%%%%%%%%%%%%%%%%%%%%%%%%%%%%%%%
%figure fscan
\begin{figure}
\begin{center}
\includegraphics[width=\linewidth,clip]{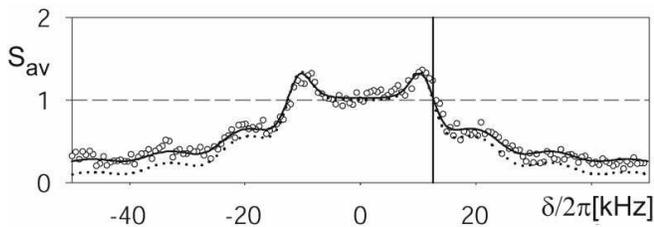}
\end{center}
\caption{Average brightness $S_{av}$ (see text) versus MS gate detuning
$\delta$. Applied gate time (75$\mu$s) is within 10\% of the ideal. Dotted
line indicates expected signal modified to include an initial temperature
$\bar{n}_s\!=\!0.3$\cite{haljan2005a}. Solid line is a fit including offset
and contrast factors to account for imperfections such as spontaneous
emission. The fit gives a sideband Rabi frequency $\eta\Omega/2\pi$=6.3kHz
and initial stretch mode temperature $\bar{n}_s$=0.3. Vertical line shows
ideal gate operation point $\delta$=$2\eta\Omega$, roughly at
$S_{av}\!=\!1$. Each point is the average of 150 PMT measurements.}
\label{fig:freqscan}
\end{figure}
%%%%%%%%%%%%%%%%%%%%%%%%%%%%%%%%%%%%%%%%%%%%%%%%%%%%%%%%%%%%%%%%%%%%%%%%%%%%%%%
%%%%%%%%%%%%%%%%%%%%%%%%%%%%%%%%%%%%%%%%%%%%%%%%%%%%%%%%%%%%%%%%%%%%%%%%%%%%
%figure tscan
\begin{figure}
\begin{center}
\includegraphics[width=\linewidth,clip]{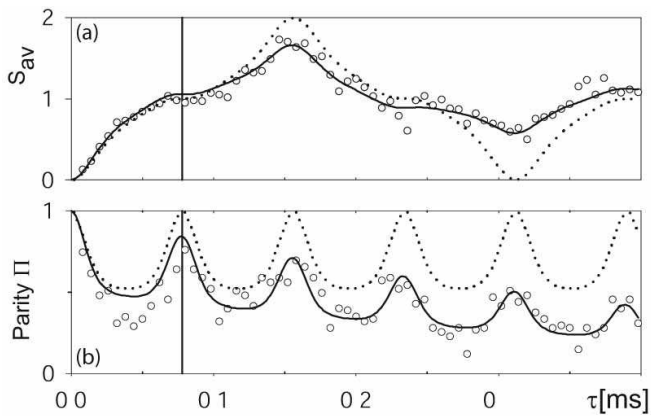}
\end{center}
\caption{Time scan of MS gate showing (a) average brightess $S_{av}$ and (b)
parity $\Pi$. Ideal gate evolution shown as dotted lines with best fit
including exponential damping shown with solid line. The fit gives a
sideband Rabi frequency $\eta\Omega/2\pi$\!=\!6.6kHz and detuning
$\delta/2\pi\!=\!12.8\textrm{kHz}\approx2\eta\Omega/2\pi$ with other
parameters the same as in figure~\ref{fig:freqscan}. Vertical line shows
gate operation time $\tau\!=\!2\pi/\delta\!\approx\!80\mu$s.}
\label{fig:timescan}
\end{figure}
%%%%%%%%%%%%%%%%%%%%%%%%%%%%%%%%%%%%%%%%%%%%%%%%%%%%%%%%%%%%%%%%%%%%%%%%%%%%%

The sequence to implement the gate begins with balancing the sideband
strengths to better than 10\% and the detunings to $\sim$100Hz. Applying the
bichromatic field for time $\tau$ to the initial state \uu~while scanning
the detuning $\delta$ pinpoints the required gate detuning to near
$2\eta\Omega$ [Fig.~\ref{fig:freqscan}]. The dynamics of the frequency scan
can be understood in terms of the evolution of entangled states of spin and
motion \cite{leibfried2003a,haljan2005a}. Assuming the initial spin state
\uu~and motional ground state $\left|n_s\!=\!0\right\rangle$, the average
ion brightness defined as
$S_{av}=2P_{\downarrow\downarrow}+P_{\uparrow\downarrow}+P_{\downarrow\uparrow}$
is $S_{av}(\tau,\delta)=1/2 (1+\cos\Phi(\tau,\delta)
e^{-|\alpha(\tau,\delta)|^2/2})$. With the detuning now fixed, the average
brightness is monitored while scanning the gate time. The time evolution
reveals the overall spin dynamics modulated by faster dynamics associated
with the phase-space evolution[Fig.~\ref{fig:timescan}(a)]. Each location of
zero slope corresponds to the ion motion returning on itself to form a closed
trajectory. The return points are most clearly visualized in the parity
signal,
$\Pi=(P_{\uparrow\uparrow}+P_{\downarrow\downarrow})-(P_{\uparrow\downarrow}+P_{\downarrow\uparrow})
=1/2(1+e^{-2|\alpha(\tau,\delta)|^2})$ [Fig.~\ref{fig:timescan}(b)]. At the
gate operation time (80$\mu$s), corresponding to the first return, the
initial state \uu~has evolved ideally to
$\Psi_1\!=\!1/\sqrt{2}(|\uparrow\uparrow\rangle + i
e^{i\phi_e}|\downarrow\downarrow\rangle)$.

\begin{figure}
\begin{center}
\includegraphics[width=\linewidth,clip]{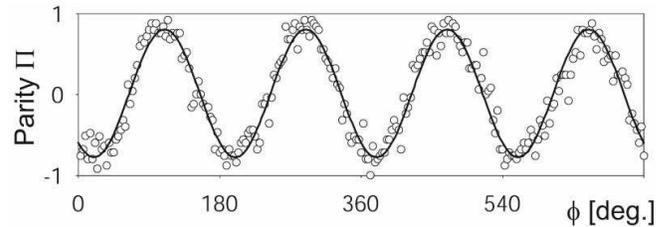}
\end{center}
\caption{Parity versus phase of analysis $\pi/2$ pulse applied to the
$\Psi_1$ state. The solid line is a sinusoidal fit yielding an amplitude
0.79(2). The fidelity of the state shown is $0.83(2)$. Each point is an
average over 50 PMT measurements and other parameters are as in text.}
\label{fig:parityscan}
\end{figure}

The simplest indicator for the quality of the entangled states formed is the
fidelity $F\!=\!\left\langle\Psi|\rho|\Psi\right\rangle$ with which the
actual density matrix $\rho$ matches the target state $\Psi$. The fidelity
for creating the Bell-like states of Eqn.~\ref{eqn:MSgate2} is simply the
sum of the two relevant diagonal population terms of $\rho$ and the
corresponding pair of off-diagonal coherences. It is easy to directly
extract the coherences for the even parity states $\Psi_{1,2}$ without
single-qubit operations. A single global $\pi/2$ analysis pulse is applied
to the state; varying the phase of the analysis pulse yields an oscillating
parity signal [Fig.~\ref{fig:parityscan}] with amplitude equal to twice the
off-diagonal coherence \cite{Bollinger1996a,sackett00}. For the case of
$\Psi_1$ as shown in Fig.~\ref{fig:parityscan}, a typical fidelity of 0.80
is achieved (which must exceed 0.5 to achieve entanglement
\cite{Bennett1996a,sackett00}).

%%%%%%%%%%%%%%%%%%%%%%%%%%%%%%%%%%%%%%%%%%%%%%%%%%%%%%%%%%%%%%%%%%%%%%%%%%%%
%figure pscan
%%%%%%%%%%%%%%%%%%%%%%%%%%%%%%%%%%%%%%%%%%%%%%%%%%%%%%%%%%%%%%%%%%%%%%%%%%%%%%%%

A full evaluation of the entangled state including a quantitative measure of
the entanglement requires access to the full density matrix, in particular
all the off-diagonal coherences. To determine the fifteen free parameters
for a normalized two-qubit density matrix requires at least as many
independent measurements. We follow closely the tomographic approach
outlined in refs.~\cite{james2001a,altepeter2004a}. The density matrix can
be decomposed in terms of a tensor product basis
$\rho\!=\!\sum_{i,j=0}^{3}{r_{ij}\sigma_i\otimes\sigma_j}$ where
$\sigma_0\equiv\mathbb{I}$, $\sigma_1\equiv\sigma_x$,
$\sigma_2\equiv\sigma_y$ and $\sigma_3\equiv\sigma_z$ are the usual
single-qubit Pauli matrices satisfying
$Tr(\sigma_i\sigma_j)\!=\!2\delta_{ij}$, and
$r_{ij}\!=\!Tr(\rho\sigma_i\otimes\sigma_j)$ are real numbers. In the
experiment we choose to perform projective measurements in the nine basis
combinations $\{\sigma_i\otimes\sigma_j, i,j\!=\!x,y,z\}$ each yielding four
possible outcomes for a total of twenty-seven independent measurements
accounting for normalization. The fluorescence measurement accesses
$\sigma_z$ projections. To implement transverse $\sigma_{x,y}$ projections,
we make use of independent single-qubit rotations to transform into the
$\sigma_z$ basis before measurement. Repeated preparation of a target state
followed by tomographic measurement is performed for 200 shots per
measurement basis. The total reconstruction time takes about 60s, dominated
by the cooling cycle and camera readout time.

A fast, direct inversion for the density matrix can be made with a minimum
complete measurement set of fifteen values $r_{ij}$. However, this process
in general leads to an unphysical density matrix due to experimental error.
Instead, maximum likelihood estimation is used to fit the data to a density
matrix form constrained to be Hermitian, normalized and positive
semidefinite. The inclusive and mutually exclusive nature of the four
measurement outcomes for each basis is taken into account by least-squares
weighting according to a multinomial distribution \cite{KendallStats}.
Systematics of the tomographic process are assessed after the fact based on
tomographic control runs of input states \uu~and \dd~assumed to be ideal.
The results from the controls are used to extract detection biases (on the
order of a few percent), microwave Rabi frequency and applied ac Stark
shifts used for qubit rotations. Statistical errors for parameters calculated
from the reconstructed density matrix are difficult to extract directly and
so are obtained using a simple numerical bootstrap method \cite{bootstrap}.
The raw shot-by-shot data are randomly resampled with replacement to generate
successive data sets from which a distribution of a parameter's value can be
obtained.

%%%%%%%%%%%%%%%%%%%%%%%%%%%%%%%%%%%%%%%%%%%%%%%%%%%%%%%%%%%%%%%%%%%%%%%%%%%
%figure_tomo
\begin{figure}
\begin{center}
\includegraphics[width=\linewidth,clip]{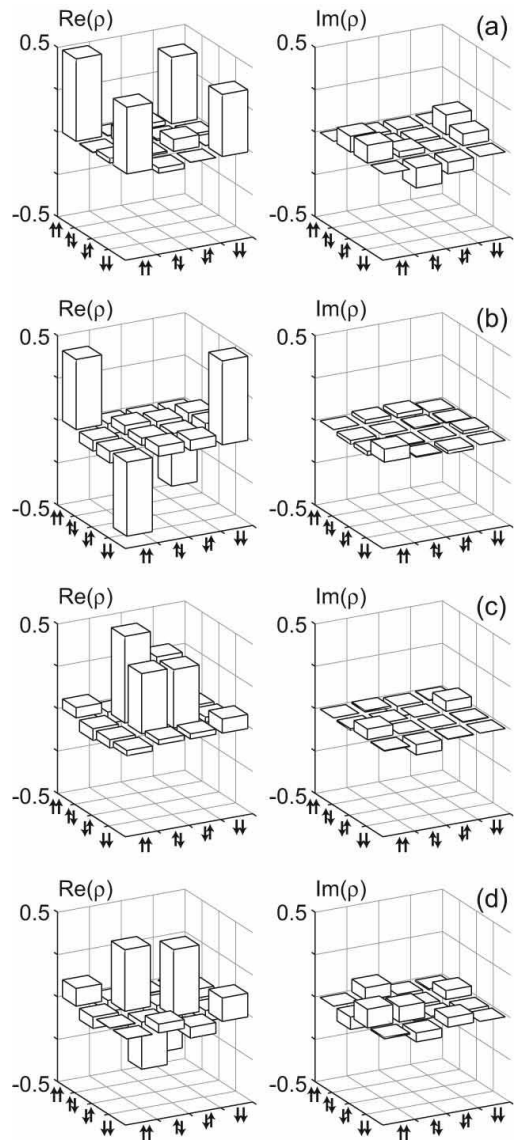}
\end{center}
\caption{Tomographically reconstructed density matrices (a)--(d) for the four
Bell-like entangled states $\Psi_1$ through $\Psi_4$ as per
Eqn.~\ref{eqn:MSgate2}. To allow direct comparison of diagonal and
off-diagonal elements, the reconstructed matrices were rotated into the real
co-ordinate using fit parameter $\phi_e\!=\!-1.1$rad for (a) and (b), and
$\phi_0\!=\!0.43$rad for (c) and (d). Each state reconstruction uses 27
independent projective camera measurements averaged over 200 runs.}
\label{fig:tomog}
\end{figure}
%%%%%%%%%%%%%%%%%%%%%%%%%%%%%%%%%%%%%%%%%%%%%%%%%%%%%%%%%%%%%%%%%%%%%%

All four Bell-like entangled states are created according to
Eqn.~\ref{eqn:MSgate2} by applying the MS gate to the different
computational states. Figure~\ref{fig:tomog} shows their reconstructed
density matrices. The inferred fidelities for the target states $\Psi_1$
through $\Psi_4$ are $F=\{$0.82(3),0.89(3),0.78(3),0.66(3)$\}$ where the
phases $\phi_e\!=\!-1.1$rad and $\phi_o\!=\!0.43$rad are considered free
parameters obtained from the fits. The tomographically obtained fidelity for
$\Psi_1$ agrees well with a simple parity-based assessment like that
discussed above. The fidelity for creating the odd-parity states
$\Psi_{3,4}$ is worse because of inaccurate preparation of the input states
\ud~and \du~($F\!\approx\!0.85$). Accounting for this factor, the fidelities
of all states are on par.

Inseparability (entanglement) of the reconstructed two-qubit states can be
tested by performing a partial transpose of the density matrix and searching
for a negative value in the resultant eigenvalue spectrum
\cite{Peres1996a,Horodecki1996a}. For example, the eigenvalue spectrum
obtained for the $\Psi_2$ case is $\{$-0.42(3),0.40(3),0.49(2),0.53(2)$\}$
compared with the ideal case $\{$-0.5,0.5,0.5,0.5$\}$. The negativity $N$
\cite{Vidal2002a,Eisert1999a}, twice the absolute value of the negative
eigenvalue, is obtained for all four target states $\Psi_i$ with values
$N\!=\!\{$0.74(6),0.84(7),0.60(5),0.42(6)$\}$. Ranging from zero for a
separable state to one for a maximally entangled one, the value gives an
indication of the degree of the entanglement. Several quantitative measures
of entanglement exist in the literature \cite{MeasureReview}, although
lacking a closed form they are in general difficult to calculate. One
standard measure that is directly calculable for two qubits is the
entanglement of formation $E_F$ \cite{Wootters1998a} again ranging from zero
for a separable state to one for a maximally entangled one. In the context of
pure states, the value of $E_F$ can be interpreted as the number $nE_F$ of
maximally entangled states required to reconstruct $n$ copies of a given
state \cite{Bennett1996a}. The experimental values for the four states shown
in Fig.~\ref{fig:tomog} are $E_F=\{$0.65(8), 0.77(9),0.49(6),0.32(6)$\}$. The
entanglement of formation is a manifestly more strict indicator for the
quality of an entangled state than the fidelity and drops quickly with
decreasing fidelity.

Among the experimental sources of gate imperfection, spontaneous emission and
fluctuating ac Stark shifts stand out as the likely primary sources of the
observed infidelity. For our setup, the gate speed
($\Omega_g\!=\!2\pi/\tau_g\!=\!2\eta\Omega$), which is proportional to the
stimulated Raman Rabi frequency $\Omega$, scales as $I\gamma^2/\Delta$ in
terms of the optical linewidth $\gamma/2\pi\!=\!60$MHz, the Raman laser
intensity $I$ and its detuning $\Delta$ (not to be confused with the
detuning $\delta$ of the net two-photon Raman transition). In addition to
generating the desired gate action, the Raman beams are responsible for a
spontaneous scattering rate $\gamma_{sc}$ per ion and residual differential
ac Stark shift $\delta\nu_{s\!t}$ of the hyperfine qubit levels. The
probability of a spontaneous photon being scattered during a gate operation
is $p_{sc}\!=\!2\gamma_{sc}\tau_g\!=\!2\beta\gamma/\Delta$ where the factor
of two accounts for the presence of two ions. A rough theoretical estimate
can be made for the prefactor
$\beta\!=\!\sqrt{2}\pi/\epsilon\zeta\eta\!\approx\!400$, which includes a
$\sqrt{2}$ factor accounting for the bichromatic field, a factor
$\epsilon\sim0.2$ characterizing the EOM Raman transition
efficiency\cite{lee2003a}, a Clebsch-Gordon related factor $\zeta\!=\!0.5$
and the Lamb-Dicke parameter $\eta\!=\!0.1$ for the stretch mode. Similarly,
the ac Stark phase acquired during a gate is
$\phi_{st}\!=\!\delta\nu_{st}\tau_g\!=\!\beta \omega_{h\!f}/\Delta$. The
relatively large value of $\omega_{h\!f}/2\pi\!=\!14.5$GHz for Cd$^{+}$,
while useful for high fidelity qubit detection, requires a significant
detuning $\Delta$ to suppress Stark shifts. Experimentally, for our modest
detuning $\Delta/2\pi\!\approx\!200$GHz we measure a value of
$\delta\nu_{st}/2\pi\!=\!75$kHz, from which we obtain $\phi_{s\!t}=12\pi$ and
infer $p_{sc}\!\approx\!0.3$. The value of $p_{sc}$ agrees roughly with the
direct theoretical estimate and predicts an infidelity $1-F\approx
0.73p_{sc}\!=\!0.2$, roughly in agreement with the observed value for
creating $\Psi_{1,2}$. The factor of 0.73 appears in the infidelity since a
spontaneous scattering event will result in a mixed state that still has
some residual overlap with the entangled target state.

Increasing the detuning $\Delta$ can reduce the relative effect of both
spontaneous emission and Stark shift (see also ref.~\cite{Haffner2003a});
however, a concomitant increase in the power of the Raman beams is required
to maintain the speed of the entangling gate (and Raman cooling), thereby
avoiding slower sources of noise such as magnetic field drift or laser
beam-steering noise. In the short term, a reasonable increase in Raman laser
detuning and power (currently $\sim$1mW) by a factor of ten would reduce the
spontaneous emission and sensitivity to ac Stark shifts by the same amount.
Ultimately detunings on the order of the large fine structure (74THz) of
Cd$^+$\cite{lee2003a} allow for significant suppression of both effects (see
also ref.~\cite{Ozeri2005a}).

In conclusion, a M{\o}lmer-S{\o}rensen gate has been realized to generate
pair-wise entanglement of clock-state ion-qubits with reduced sensitivity to
interferometric phase fluctuations of the Raman beams. The tomographic
reconstruction used to assess the resultant entangled states demonstrates
universal two-qubit control, which is being directly applied to investigate
prototype quantum algorithms \cite{groverTBP}.

This work is supported by the National Security Agency and Advanced Research
and Development Activity under Army Research Office contract
W911NF-04-1-0234, and the National Science Foundation Information Technology
Research program.
%\begin{acknowledgments}
%\end{acknowledgments}

\bibliographystyle{apsrevmod}
%\bibliography{ion,ion2}% Produces the bibliography via BibTeX.

\begin{thebibliography}{43}
\bibitem[{pha()}]{phasenote}
\bibinfo{note}{The relative phases between the entangled states are chosen for
  simplicity since they do not affect the tomography.}
\end{thebibliography}

\end{document}